\documentclass[twocolumn]{openjournal}

\usepackage{xcolor}
\usepackage{textgreek}
\usepackage[utf8]{inputenc}
\usepackage[english]{babel}

\usepackage{hyperref}
\hypersetup{
    unicode, 
    colorlinks=true,
    linkcolor=linkcolor,
    citecolor=linkcolor,
    filecolor=linkcolor,
    urlcolor=linkcolor,
}
\usepackage{color,colortbl}
\definecolor{linkcolor}{rgb}{0.0,0.3,0.5}
\usepackage{tensind}
\tensordelimiter{?}
\DeclareGraphicsExtensions{.bmp,.png,.jpg,.pdf}
\usepackage{verbatim}
\usepackage[normalem]{ulem}
\usepackage{orcidlink}
\usepackage{soul}
\usepackage{amsmath}

\addto\extrasenglish{%

}

\newcommand{\criptic}{\texttt{CRIPTIC}}
\newcommand{\gizmo}{\texttt{GIZMO}}
\newcommand{\congruents}{\texttt{CONGRuENTS}}
\newcommand{\slug}{\texttt{SLUG}}
\defcitealias{shah25}{S25}
\defcitealias{criptic}{K22}

\graphicspath{ {./figs/} }

\begin{document}
\title{Spirited Away: Advective loss of cosmic rays into the Milky Way's circumgalactic medium explains the Large Magellanic Cloud's low $\gamma$-ray luminosity}

\author{Taaseen Islam \orcidlink{0009-0008-2314-9904}}
\email{taaseen.islam@adelaide.edu.au}
\affiliation{Research School of Astronomy \& Astrophysics, Australian National University, 233 Mt Stromlo Rd, Stromlo, ACT 2611, Australia}
\affiliation{Centre for the Subatomic Structure of Matter, Adelaide University, Adelaide, SA 5005, Australia}

\author{Hilay Shah\orcidlink{0000-0002-9136-6731}}
\email{Hilay.Shah@anu.edu.au}
\affiliation{Research School of Astronomy \& Astrophysics, Australian National University, 233 Mt Stromlo Rd, Stromlo, ACT 2611, Australia}

\author{Mark R. Krumholz\orcidlink{0000-0003-3893-854X}}
\email{mark.krumholz@anu.edu.au}
\affiliation{Research School of Astronomy \& Astrophysics, Australian National University, 233 Mt Stromlo Rd, Stromlo, ACT 2611, Australia}

\author{Roland Crocker\orcidlink{0000-0002-2036-2426}}
\email{rcrocker@fastmail.fm}
\affiliation{Research School of Astronomy \& Astrophysics, Australian National University, 233 Mt Stromlo Rd, Stromlo, ACT 2611, Australia}

\begin{abstract}
Models of galactic-scale cosmic ray production and transport have successfully reproduced the radio and $\gamma$-ray spectra of many galaxies; however, one notable exception is the Large Magellanic Cloud (LMC), where models that successfully fit other galaxies consistently overestimate its $\gamma$-ray flux. Here we investigate this discrepancy by applying the \criptic~cosmic ray transport code to recent magnetohydrodynamic simulations of the LMC's interactions with the Milky Way circumgalactic medium that accurately reproduce observations of the LMC's magnetic field structure. We recover a simulated $\gamma$-ray luminosity that is close to the observed luminosity for a wide range of cosmic ray transport models, while our control simulations of an isolated LMC not interacting with the Milky Way show the same over-prediction problem as previous investigations. Simulations including the CGM interaction yield lower $\gamma$-ray luminosities because in them cosmic rays escape the galaxy primarily via advection, significantly decreasing the emission from collisional processes and changing the dominant $\gamma$-ray emission mechanism from pion decay to inverse Compton emission. Comparisons of the detailed spectral shape show that our interacting LMC models match the observed $\gamma$-ray spectrum at photon energies below $\sim 10$ GeV, but still slightly overestimate the flux at higher energies, suggesting either more strongly energy-dependent transport than the models we have explored, or a cosmic ray injection spectrum steeper than the $p^{-2.2}$ that we adopt.\\
\end{abstract}


\maketitle

\section{Introduction}
\label{sec:intro}

Cosmic rays (CRs) are a critical constituent of the interstellar medium (ISM). The ionization they provide drives chemistry in regions too dense and well-shielded for photons to penetrate \citep{Padovani20a, Gabici22a}, the forces they exert may be critical for driving galactic winds and regulating star formation \citep[e.g.,][]{Hopkins20a, Crocker21, Crocker21b, RP23}, and their interactions with thermal gas, magnetic fields, and starlight in the ISM drive the majority of galaxies' diffuse non-thermal emission \citep[e.g.,][]{blumenthal_bremsstrahlung_1970, kelner06}. However, our ability to understand and predict all of these processes is limited by our lack of a consistent theory for CR transport through the ISM. Because CRs are charged, they can be deflected by ISM magnetic fields, and because the flow of CRs represents a current, the CRs themselves can influence those magnetic fields, leading to complex and poorly-understood self-interactions \citep[e.g.,][]{Zweibel13a, Zweibel17a, Hopkins25}. A fully satisfactory macroscopic theory for the transport of CRs produced by all these effects has thus far eluded the community \citep[e.g.,][]{Hopkins22c, Kempski22a}.

Given this situation, one way to make progress is to use the non-thermal emission produced by CR-ISM interactions as a diagnostic for transport, since different transport theories lead to different predictions for the spectra and morphologies of the resulting emission. The strategy has been applied across a range of scales, from individual spatially-resolved sources within the Milky Way \citep[e.g.][]{Thomas20a, Recchia21a, criptic2}, to the Milky Way as a whole \citep[e.g.,][]{Johannesson16a}, to external galaxies \citep[e.g.,][]{Hopkins21a, werhahn21b, werhahn21c, Ambrosone22a, roth23, Roth24a} to the entire diffuse $\gamma$-ray sky \citep[e.g.,][]{roth21, Owen22a, Ambrosone24a}. The Large Magellanic Cloud (LMC) is a convenient observational target for this approach due to its proximity and large angular size. In particular, observations of the LMC's radio and $\gamma$-ray non-thermal emission \citep[e.g.,][]{ajello20, fermi43} as well as measurements of Faraday rotation \citep[e.g.,][]{gaensler05, livingston_24_LMCRM} have provided constraints on both the LMC's CR population and its magnetic field structure. However, despite these high-quality data, existing CR transport models for the LMC have consistently overestimated its $\gamma$-ray luminosity compared to observations. This includes both models based on direct numerical simulations \citep{bustard20, werhahn21b} and those based on semi-analytic models \citep{roth23, Roth24a}. 

This does not appear to be a general failing of the models: the same models successfully predict the empirical far-infrared (FIR)-$\gamma$ ray correlation of galaxies \citep{kornecki20}, and make reasonable predictions for the Small Magellanic Cloud (SMC)\footnote{We note there are significant statistical deviations between different measurements of the SMC's $\gamma$-ray spectrum, as discussed by \citet{roth23}, who compare data from \citet{ajello20} and \citet{fermi43}. As a result of these discrepancies, it is possible to say that current models yield qualitatively reasonable predictions for the SMC's total $\gamma$-ray luminosity \citep[e.g.][]{kornecki20, werhahn21b}, but not whether they reproduce its $\gamma$-ray spectrum.}. The problem appears to be that the LMC itself is an outlier from the trend on which other galaxies sit, and which these models successfully reproduce.

Several explanations of the LMC's $\gamma$-ray dimness have been proposed. \citet{kornecki20} suggest that a significant fraction of CR protons escape from the LMC because of its interaction with the Milky Way's circumgalactic medium (CGM). The LMC is currently falling into the Milky Way, causing ram pressure stripping of the LMC's ISM and possibly its CRs as well. \citet{roth23} propose that because there has been a recent sharp increase in the LMC's star formation rate, the LMC's CR population may not yet have equilibrated to match its higher star formation rate. However, neither these nor any other hypotheses have been tested in detail, leaving our closest neighbor---a system that should in principle be an ideal laboratory for testing CR propagation models---an embarrassing anomaly.

The purpose of this work is to investigate CR transport models in the LMC to determine a cause for its low $\gamma$-ray luminosity. To achieve this we will apply the CR propagation software \criptic~\citep[hereafter \citetalias{criptic}]{criptic} to recent magnetohydrodynamic simulations of the LMC that include the effects of its interaction with the Milky Way's CGM, and that have been calibrated against observational constraints of the LMC's magnetic field strength and structure. \autoref{sec:methods} describes our method in detail, while \autoref{sec:results} then evaluates the results of the simulations, and \autoref{sec:discussion} discusses the implications of our results.

\section{Methods}
\label{sec:methods}

We calculate CR propagation and non-thermal emission in the LMC by applying the \criptic~CR transport and emission code to simulations of the LMC. \autoref{subsec:criptic} describes our method for using \criptic~to calculate the CR distribution in the LMC, \autoref{subsec:spectrum} explains how we compute the spectrum these CRs emit, and \autoref{subsec:simulations} describes the set of simulations we run.

\subsection{Method for \criptic~postprocessing}
\label{subsec:criptic}

The \criptic~code uses an It\^o Calculus approach to solve the Fokker-Planck equation for CR propagation in either the pitch angle-averaged \citepalias{criptic} or pitch angle-dependent \citep{criptic2} formulation for an arbitrary background state of thermal gas, magnetic fields, and radiation fields. It includes all of the dominant loss and radiative emission processes for CRs---nuclear inelastic scattering by hadrons \citep{kafexhiu16}, inverse Compton scattering, synchrotron emission, and bremsstrahlung by leptons \citep{blumenthal_bremsstrahlung_1970}, Coulomb and ionization losses of all particles, and annihilation of positrons---and includes a robust treatment of secondary particle production. We refer readers to \citetalias{criptic} for a detailed description of the methods, and instead here we focus on the ingredients required for our application of \criptic: descriptions of the background gas, magnetic field, and radiation field through which the CRs propagate, coupled with a model for sources of CRs.

\subsubsection{The LMC simulations}
\label{subsec:lmc_sims}

\begin{figure*}
    \includegraphics[width=\textwidth]{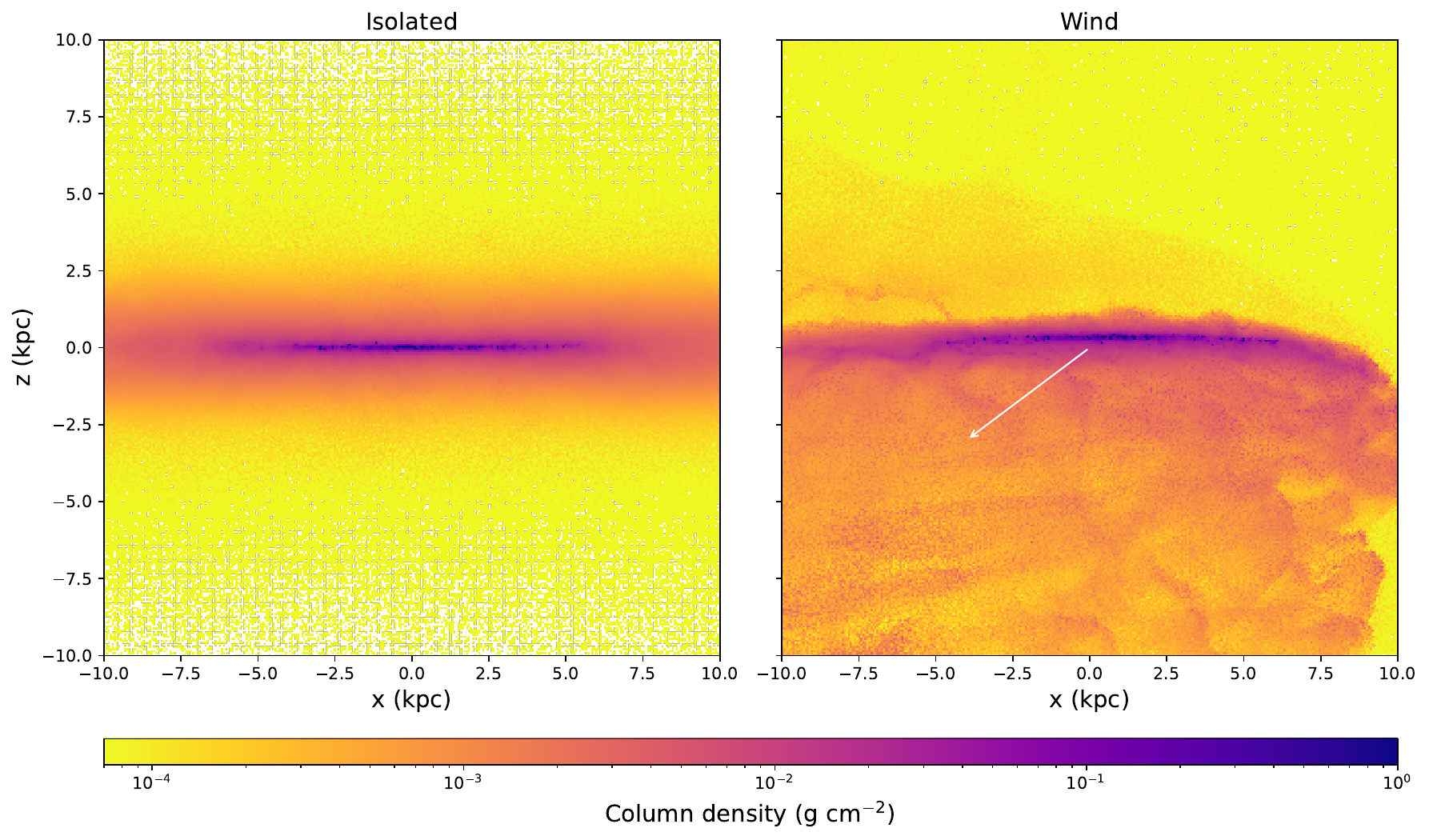}
    \caption{Snapshots from the two LMC simulations from \citetalias{shah25} that we use in this work at their final times, after 500 Myr of simulation time. The simulations have been translated and rotated such that the galaxy centers are at rest at the origin and the disk's angular momentum is aligned with the $z$-axis. Colors show column density integrated over a $40$ kpc-thick slab centered on the galaxy center. The left panel shows the ``isolated'' simulation \texttt{I-2-6-M}, while the right shows the ``wind'' simulation \texttt{W-2-6-M}, which includes in the Milky Way CGM (equivalent to a wind in the LMC's rest frame). The white arrow in the right panel shows the direction of the wind in the LMC rest frame.\\}
    \label{fig:lmc_sim}
\end{figure*}

Our background gas and magnetic field states are taken from two simulations of the LMC by \citet[hereafter \citetalias{shah25}]{shah25}. These authors carried out a suite of ideal magnetohydrodynamic simulations of LMC-like galaxies using \gizmo~\citep{meshoid, hopkins16a}; in this suite they varied the initial magnetic field configuration in order to attempt to reproduce observed Faraday rotation measure statistics from \citet{gaensler05}. They also studied the effects of the LMC's interaction with the Milky Way CGM by carrying out pairs of simulations that differ only in that one is for an isolated LMC-like galaxy, while the other includes a background gas through which the LMC-like galaxy travels with a density distribution chosen to match that experienced by the LMC during its approach to the Milky Way. For the purposes of this work, we use \citetalias{shah25}'s simulation \texttt{W-2-6-M}, which includes the CGM and provides an excellent match to the Faraday rotation measure data, and \texttt{I-2-6-M}, which is its control counterpart without a CGM. We show snapshots from these two simulations in \autoref{fig:lmc_sim}.

Both of these simulations ran for a total of 500 Myr, with the CGM density profile in \texttt{W-2-6-M} set so that at $t=500$ Myr the CGM conditions match those being experienced by the LMC today. For the purposes of our calculations here, we make use of the last 40 Myr of the two simulations, i.e., the period from $t=460-500$ Myr, with snapshots of the simulation state at intervals of 0.25 Myr over this period. We post-process these snapshots to produce realistic abundances for free electrons and the various possible ionization states of H and He using the procedure described in Section 2.4 of \citetalias{shah25}. This provides us with estimates of the abundances of H, H$^+$, H$_2$, He, He$^+$, He$^{++}$, and $e^{-}$ at all points in the simulation; these are required by \criptic, since its models for ionization, Coulomb losses, and bremsstrahlung distinguish between these different ISM species.

To use the simulations to provide a gas and magnetic field background for \criptic, at any given time $t$ in our \criptic~simulations we hold in memory the pair of \citetalias{shah25} simulation snapshots for which $t_i < t < t_{i+1}$, where $t_i$ is the time of the $i$th snapshot. Any time we require either the total gas density, the density of particular chemical species, or the magnetic field at a position $\mathbf{x}$ and time $t$, we compute them by evaluating \gizmo's cubic spline kernel (see \citealt{meshoid}) at that position for both snapshots, then linearly interpolate in time between them; we organize the snapshot data into a kd-tree structure to keep the cost of this evaluation manageable \citep{kdtree}. In order to minimize interpolation errors, before carrying out this procedure we shift the positions and velocities of all particles in the snapshots from \texttt{W-2-6-M}---in which the LMC moves through the CGM---into a reference frame comoving with the centre of the LMC.

\subsubsection{Radiation fields} \label{subsubsec:radfields}

The \citetalias{shah25} simulations we use do not model the radiation field around the LMC, but a model for radiation fields is necessary for calculation of inverse Compton scattering---a key emission process for CR electrons. We must therefore add a model for radiation fields to supplement the LMC models from the \gizmo~simulations. Radiation fields in \criptic~are described as a sum of dilute blackbodies characterized by temperatures $T_\mathrm{BB}$ and dilution factors $W_\mathrm{BB}$ (defined as the ratio of the radiation energy density in that field to that of a blackbody field with equal temperature). We construct our estimates of these quantities within the galaxy following the approach used in the \congruents~code \citep[see][their Section 2.3]{roth23}. \congruents~uses a radiation field with six components: a blackbody cosmic microwave background (CMB) component with temperature $T_\mathrm{CMB}=2.73$ K; a modified blackbody dust component for far-infrared thermal dust emission (which we treat as a true blackbody, since \criptic~does not support modified blackbodies); three blackbody components with temperatures $3000$ K, $4000$ K, and $7500$ K representing cool, warm, and hot stars; and a far-UV component. We omit the far-UV component because its spectral shape is not supported by \criptic, and because it makes only a very small contribution to the total energy budget ($\approx 3.5\%$ in the Solar neighborhood---\citealt{Draine}).

The dust component's temperature is given by an empirical fit taken from \citet{magnelli_14}, which depends on the star formation rate and stellar mass of the galaxy. The dilution factors of the dust component and the three stellar components are derived from an empirical fit calibrated against DustPedia galaxies, which also depends on star formation rate and stellar mass \citep{dustpedia}. We evaluate these fits using the observed present-day LMC star formation rate $\dot{M}_* \approx 0.2$ M$_\odot$ yr$^{-1}$ \citep{harris_zaritsky_09} and stellar mass $M_* = 2.7\times 10^9$ M$_\odot$ \citep{LMC_mass}. This yields a dust temperature $T = 28.1$ K, and a set of dilution factors $W_0$ and corresponding energy densities $U_0 = W_0 a_R T^4$ for the radiation field components that we list in \autoref{tab:radfield}.

\begin{table}
    \centering
    \begin{tabular}{lccc}
        \hline\hline
        \\[-2ex]
        Component & $T$& $\log W_0$ & $U_0$\\ 
        & [K] & & [eV cm$^{-3}$] \\
        \\[-2ex]
        \hline
        \\[-2ex]
        CMB  & 2.73 & 0 & 0.26\\
        Dust &  28.1 & $-4.40$ & 0.12\\
        Cool stars & 3000  & $-13.2$ & 0.025 \\
        Warm stars & 4000 & $-13.8$ & 0.019 \\
        Hot stars & 7500 & $-14.5$ & 0.045 \\ 
        \hline
        \\[-2ex]
        Total & & & 0.46 \\
        \hline\hline
        \end{tabular}
        \caption{Radiation field temperatures, central dilution factors, and central energy densities.
        \label{tab:radfield}
        }
    \end{table}

For all components other than the CMB, the dilution factors $W_0$ that we compute from this procedure are intended to be those within the LMC disk. Since CRs may escape outside the galaxy over the course of a \criptic~simulation, we also require a model for how the radiation field falls off as we move away from the disk. For this purpose, we adopt a simple empirical model 
\begin{equation} \label{eq:W_of_r}
    W(\mathbf{r}) = \frac{2W_0}{\left(1+r/R_0\right)^2 + \left(1+|z|/Z_0\right)^2},
\end{equation}
where $r$ and $z$ are the radius and height in a cylindrical coordinate system centered on the galactic center and with the galactic disk lying in the plane $z=0$, and $R_0 = 4.8$ kpc and $Z_0 = 0.48$ kpc are the scale radius and height of the LMC (matching the values used by \citetalias{shah25}, which in turn are taken from \citealt{Lucchini20a}). We choose this functional form because it has the desired limiting behavior, i.e., $W(\mathbf{r})\to W_0$ as $|\mathbf{r}|\to 0$ and $W(\mathbf{r})\propto 1/|\mathbf{r}|^2$ as $|\mathbf{r}|\to\infty$, and because it gives the radiation field a shape that reflects that of the underlying disk.

\subsubsection{Cosmic ray injection} \label{subsec:cr_inj}

The second part of our post-processing pipeline is a recipe for CR injection. In a \criptic~simulation, a source of CRs is characterized by a position, velocity, and luminosity, and thus we must have a recipe to decide how to place and assign properties for these sources. We assume supernovae (SNe) are the only significant CR accelerators in the LMC, and thus the sources correspond to the times and locations of SNe in the \citetalias{shah25} simulations. In the simulations SNe are handled using the \slug~stochastic stellar population synthesis code \citep{SLUG, Krumholz15b}, which for each ``star particle'' that forms in the simulations draws a corresponding stellar population from the initial mass function and uses that draw together with stellar evolutionary tracks to determine the times of SNe---see \citetalias{shah25} for details.

In practice, this means that each snapshot contains a set of star particles, each of which records how many SNe it has produced up to that snapshot, thereby encoding the locations and times of these SNe. We therefore proceed as follows: we first generate a list of SN positions and times by identifying all star particles that have produced SNe between any given pair of snapshots. For each such SN we then assign an exact time, position, and velocity by drawing from a uniform distribution between the times and particle positions and velocities at the two snapshots. To minimize numerical complications, turn-on times are rounded to the nearest 10 kyr.

For each source we create, we assign a luminosity $L_{\mathrm{src},p} = 3.17 \times 10^{37} \text{ erg s}^{-1} $ in CR protons and $L_{\mathrm{src},e} = 3.17 \times 10^{36} \text{ erg s}^{-1} $ in CR electrons, and leave the source on for a time $t_\mathrm{src}=0.1 \text{ Myr}$. This gives a total energy of $10^{50} \text{ erg}$ injected into CR protons and $10^{49} \text{ erg}$ injected into primary electrons over the source lifetime. We assign the injected particles---both protons and electrons---a spectrum of momenta $dn/dp\propto p^{-2.2}$ over an energy range from $0.1-10^6\text{ GeV}$, consistent with observations of $\gamma$-ray bright supernova remnants \citep{damiano_11}.

\subsection{Calculation of the emitted spectrum}
\label{subsec:spectrum}

As discussed in \citetalias{criptic}, \criptic~calculates photon emission from all the major processes by which CRs emit, including hadronic collisions leading to pion production, bremsstrahlung, inverse Compton, synchrotron, and annihilation-in-flight of positrons. The code calculates emission via these processes at user-specified photon energies or frequencies. For the purposes of this work, we perform this calculation at two sets of photon energies, one in the $\gamma$-ray band and one in the radio band. For the former, we calculate emission at $\varepsilon = 0.01,0.1,0.17,0.55,1.75,5.47,17.31,94.97,316.23,$ and $1000$ GeV, while for the latter we compute at photon energies $\varepsilon = 2\times 10^{-7}$, $3.14\times 10^{-7}$, $6.20\times 10^{-7}$, $6.69\times 10^{-7}$, $9.39\times 10^{-7}$, $10^{-6}$, $6.31\times 10^{-6}$, $2\times 10^{-5}$, and $10^{-4}$ eV (corresponding to photon frequencies $\nu = 0.048, 0.076, 0.150, 0.162, 0.227, 0.242, 1.53, 4.84,$ and 24.2 GHz). Our middle six $\gamma$-ray energies correspond to the central energies of the bins over which the LMC $\gamma$-ray spectrum is reported in the 4FGL-DR4 catalog \citep{fermiDR4}, the second lowest and highest energies correspond to the edges of the lowest- and highest-energy catalog bins, and the lowest and highest energies we compute extend our predictions somewhat outside the Fermi detection range. Similarly, our chosen radio frequencies correspond to the values provided in the compilation of \citet{For18a}.

The calculation of radio emission requires some additional consideration, for two reasons. First, over the $\sim 0.05 - 25$ GHz frequency range of interest, synchrotron emission from CRs is often overlapping with free-free emission, which \criptic~does not model. Thus testing our CR model against observed radio emission requires a model for free-free emission. For this work, we describe the free-free spectrum as a power law in frequency as $F_\nu \propto \nu^{-0.118}$ \citep[Equation 10.7]{Draine}. To normalize the spectrum against observed radio data, we assume that the highest frequency data point we use (at 24.2 GHz) is dominated by free-free emission, so we normalize the free-free spectrum such that this data point is perfectly fit by the modeled free-free and simulated synchrotron emission. This leaves the lower frequency data points as comparison points which our simulated synchrotron emission would ideally fit. 

The second consideration is that, for the lower end of our frequency range of interest, electrons have long cooling timescales. The Lorentz factor $\gamma_e$ for electrons with cutoff frequency $\nu_c$ is radiating in a magnetic field of strength $B$ is
\begin{equation}
    \gamma_e \approx \sqrt{\frac{2\pi m_e c \nu_c}{e B}},
\end{equation}
and if these electrons cool via inverse Compton and synchrotron radiation in an environment with a total magnetic plus radiation energy density $U$, the cooling time is
\begin{equation}
\label{eq:cooling_time}
    t_\mathrm{c} \approx \frac{3 m_e c}{4 \gamma_e \sigma_T U} = 32 B_{-6}^{1/2} U_0^{-1} \nu_9^{-1/2}\, \mathrm{Myr},
\end{equation}
where $B_0 = B / 1\,\mu$G, $U_0 = U/1$ eV cm$^{-3}$, and $\nu_9 = \nu / 1$ GHz. The LMC simulations of \citetalias{shah25} that we use have mean magnetic field strengths $B\approx 2$ $\mu$G in the galactic disk, and the typical radiation plus magnetic energy density in the disk (using the radiation field figures from \autoref{tab:radfield}) $U \approx 0.56$ eV cm$^{-3}$, so inserting those figures we have $t_\mathrm{c}\sim 300$ Myr for our lowest frequencies at $\nu \approx 0.1$ GHz. As discussed in \autoref{subsec:simulations}, this is significantly longer than the time for which we run our simulations, which is set to ensure that the protons responsible for producing $\gamma$-rays reach steady-state. Consequently we must regard the synchrotron luminosities predicted by our simulations as lower limits on the true synchrotron luminosity, since if cooling (rather than escape) is the main loss mechanism for electrons, then our simulations will not run long enough for the synchrotron-producing electron population to have reached steady-state.

\subsection{Propagation models and simulations}
\label{subsec:simulations}

The final ingredient required by \criptic~to calculate CR transport and non-thermal emission is a model for CR propagation. In a \criptic~calculation CRs are always advected along with the gas flow, but on top of this there can be a range of processes that cause CRs to flow relative to the gas. In \criptic's pitch angle-averaged formalism, which we use here, these processes are parameterized by diffusion coefficients parallel and perpendicular to the magnetic field and streaming velocities along the field, all of which can be arbitrary functions of CR momentum and background gas state. For this work we focus on transport by diffusion parallel to magnetic field lines, with a diffusion coefficient that depends on particle momentum $p$ as
\begin{equation} \label{eq:D_powerlaw}
    D_{\parallel} = D_{\parallel, 0}\left(\frac{p}{m_pc}\right)^q
\end{equation}
where $m_p$ is the proton mass, and $D_{\parallel, 0}$ and $q$ are the diffusion coefficient at a CR momentum of $m_p c$ and an index describing the variation of diffusion rate with momentum; we vary both of these parameters from run to run. We also enforce that $D_\parallel$ cannot fall below the Bohm limit\footnote{In principle, the Bohm limit should also apply to perpendicular diffusion, but parallel diffusion will almost always dominate. Including perpendicular diffusion within \criptic~is possible, but it would induce a relatively high computational cost for a small increase in accuracy. We therefore choose to neglect perpendicular diffusion and set $D_\perp=0$.} \citep{bohmlimit},
\begin{equation}\label{eq:Bohm}
    D_{\parallel, \rm Bohm}=\frac{1}{3}r_Gv=\frac{1}{3}\frac{pvc\sin(\theta)}{e|\mathbf{B}|} \approx \frac{\pi}{12}\frac{pvc}{e|\mathbf{B}|}
\end{equation}
where $r_G=pc\sin(\theta)/e|\mathbf{B}|$ is the particle gyroradius, $v$ is the particle speed, $e$ is the electron charge, $|\mathbf{B}|$ is the magnetic field strength, and in the final step we have averaged over a uniform distribution of pitch angle cosines $\cos \theta$. Finally, to avoid prohibitively small time steps we enforce an absolute maximum diffusion coefficient of $10^{32}$ cm$^2$ s$^{-1}$ on all CRs; this limit only becomes relevant for a tiny fraction of particles with very high energies propagating through regions of very weak magnetic field---$(pc/\mathrm{GeV}) (B/\mu\mathrm{G})^{-1} \gtrsim 10^{10}$---and thus has negligible effects on the results.

We summarize the set of simulations we have carried out in \autoref{tab:RunLabels}.  For each choice of transport model, we carry out two simulations, one for the ``isolated'' and one for the ``wind'' case as described in \autoref{subsec:lmc_sims}. We denote runs as \texttt{iDXXQYY} or \texttt{wDXXQYY}, where \texttt{i} and \texttt{w} denote a wind and isolated simulation respectively, \texttt{XX} is the value of $\log D_{\parallel,0}$ in cgs units and \texttt{YY} is the value of $q$. Our range of $D_{\parallel,0}$ values, from $10^{27} - 10^{29}$ cm$^2$ s$^{-1}$, covers the range generally inferred for GeV-energy CRs in the Milky Way \citep[e.g.][]{Johannesson16a}, as well as the values used by \citet{werhahn21b} and \citet{roth23} to model the LMC. Our choices of $q = 0$ and $q=1/3$ correspond to the simplest possible model, and to a slightly more complex one where the energy-dependence of the diffusion coefficient follows the scaling expected for confinement by extrinsic turbulence following a Kolmogorov cascade \citep{berezinsky90}.

All simulations are run for 40 Myr, as noted in \autoref{subsec:lmc_sims}, and use a packet injection rate $\Gamma = 6\times 10^{-10}$ s$^{-1}$, a step size parameter $c_\mathrm{step} = 0.25$, and a secondary sampling factor $f_\mathrm{sec} = 0.2$. See \citetalias{criptic} for a description of these parameters and their meanings.

\begin{table}
    \centering
    \begin{tabular}{lccc}
        \hline\hline
        \\[-2ex]
        Name & Wind & $\log D_{\parallel,0}$ & $q$\\
        & & [cm$^{-2}$ s$^{-1}$] \\
        \\[-2ex]
        \hline
        \\[-2ex]
        \texttt{iD27Q0} & N & 27 & 0  \\
        \texttt{iD27Q1/3} &N & 27 & 1/3  \\
        \texttt{iD28Q0} & N &28 & 0  \\
        \texttt{iD28Q1/3} &N & 28 & 1/3 \\
        \texttt{iD29Q0} & N &29 & 0 \\        
        \texttt{iD29Q1/3} & N &29 & 1/3 \\        
        \texttt{wD27Q0} & Y & 27 & 0 \\
        \texttt{wD27Q1/3} & Y & 27 & 1/3 \\
        \texttt{wD28Q0} & Y & 28 & 0 \\
        \texttt{wD28Q1/3} & Y & 28 & 1/3 \\
        \texttt{wD29Q0} & Y & 29 & 0 \\
        \texttt{wD29Q1/3} & Y & 29 & 1/3 \\
        \\[-2ex]
        \hline\hline
    \end{tabular}
    \caption{Summary of CR propagation models tested. See equation \eqref{eq:D_powerlaw} for the meaning of $D_{\parallel, 0}$ and $q$.\\
    \label{tab:RunLabels}
    }
\end{table}

\section{Results}
\label{sec:results}

We begin our presentation of results in \autoref{subsec:onecase} with a case study of two matched runs---one isolated and one wind---to introduce the qualitative outcomes of the simulations. We then examine how the results vary with diffusion coefficient in \autoref{subsec:diffusion_on_emission}.

\subsection{Case study}
\label{subsec:onecase}

\begin{figure*}
        \centering
    	\includegraphics[width=0.98\linewidth]{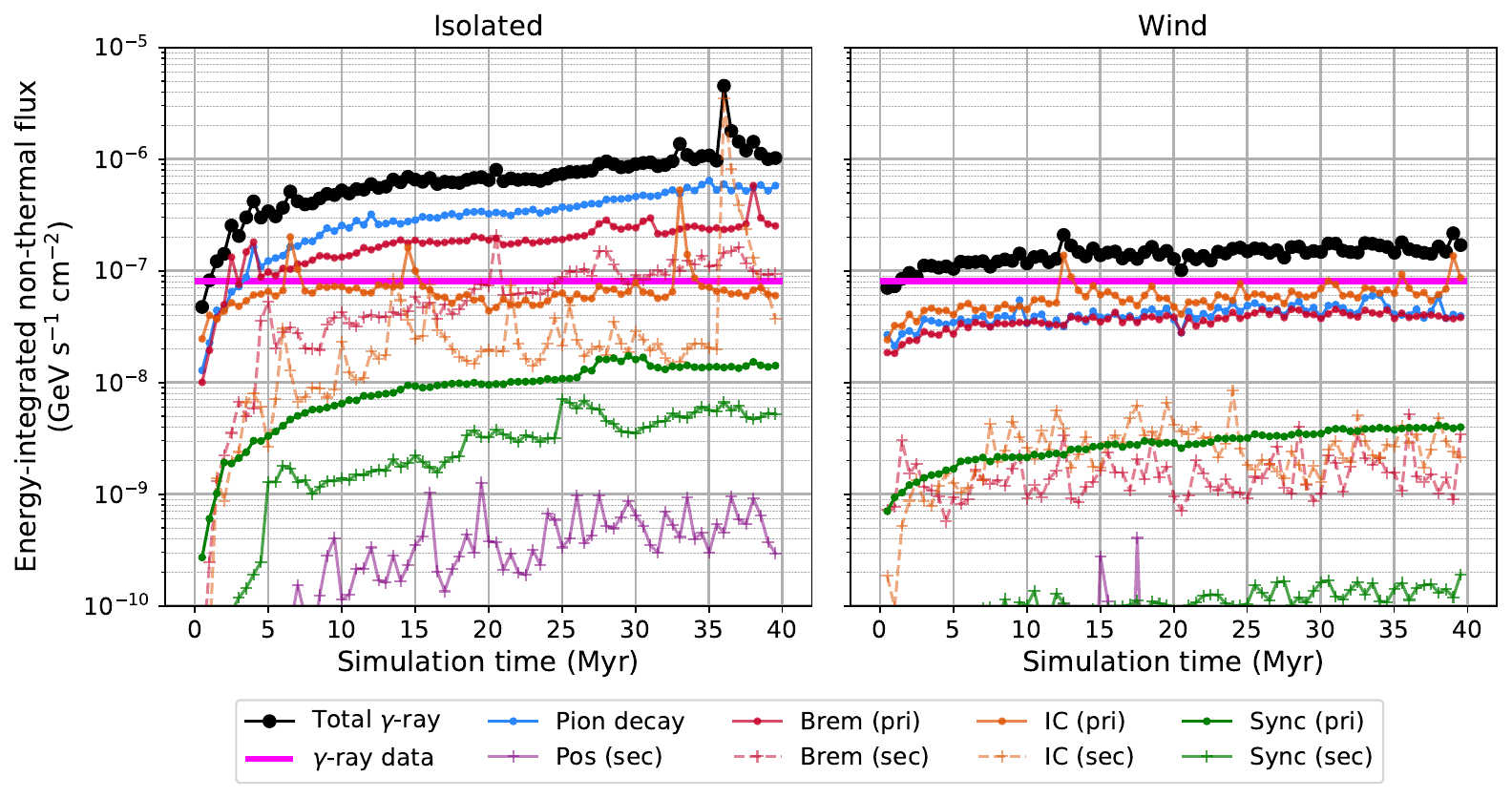}
    	\caption{Energy-integrated non-thermal flux (multi band) as a function of time in the \texttt{iD28Q0} (left) and \texttt{wD28Q0} (right) simulations, where $t=0$ corresponds to when we start following CR propagation and $t=40$ Myr to the state of the LMC as it exists now. Black points show the total flux from all processes, while colored lines with points show the contribution from each emission process---pion decay, bremsstrahlung, inverse Compton emission, synchrotron emission, and in-flight annihilation of positrons; radiation produced by leptons is also separated into emission from primary electrons and secondary leptons. The pink solid line shows the integrated $\gamma$-ray flux determined from the observed spectrum reported in the Fermi 4FGL-DR4 catalog \citep{fermiDR4}. Simulated $\gamma$-ray fluxes are integrated over the bounds of the Fermi data ($0.17-94.97$ GeV), and synchrotron fluxes are integrated over the total range of the points we show in \autoref{fig:spectra_case} ($2\times 10^{-7} - 10^{-4}$ eV).}
        \label{fig:timeseries_case}
\end{figure*}

\begin{figure*}
        \centering
    	\includegraphics[width=0.98\linewidth]{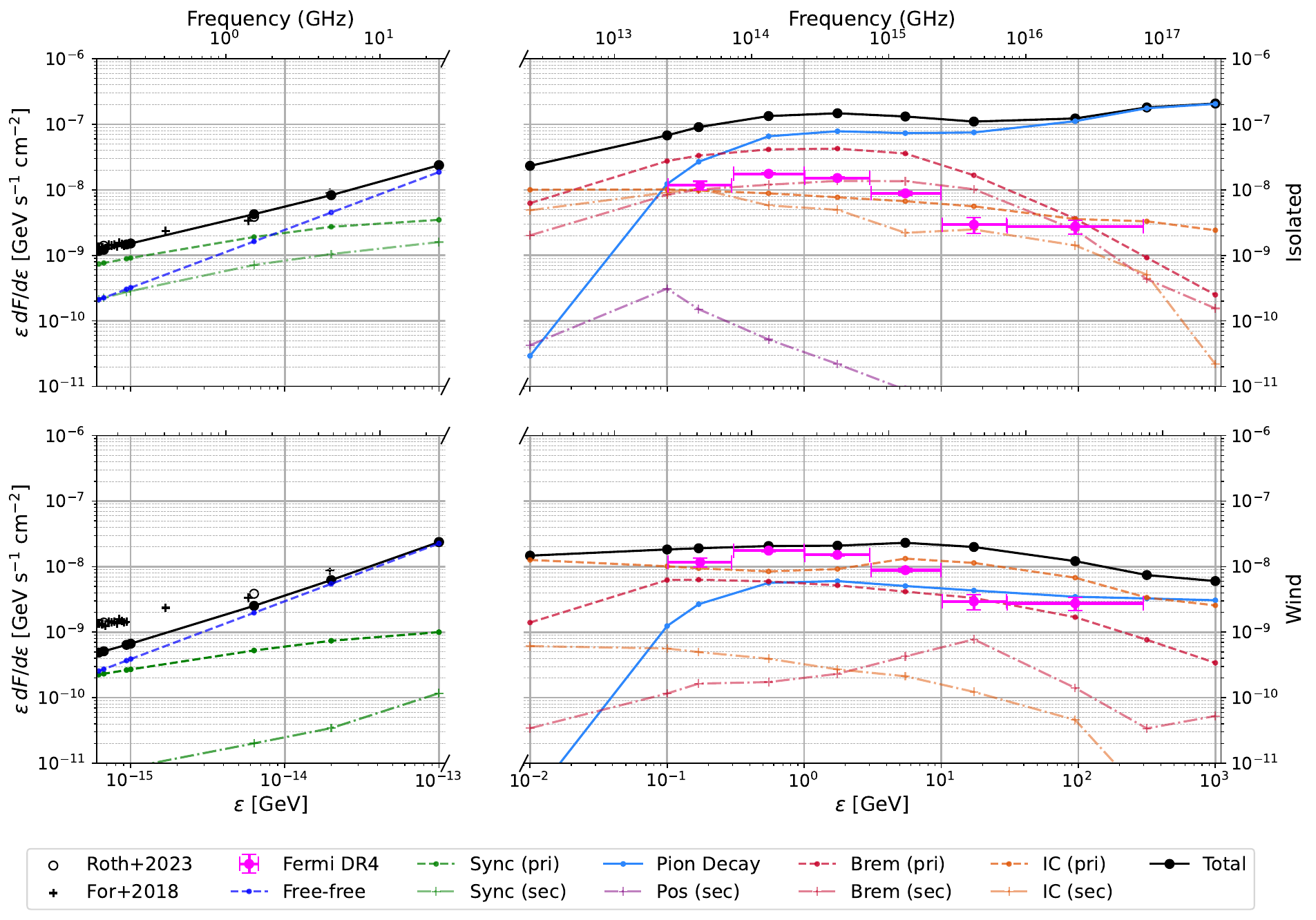}
    	\caption{Radio and $\gamma$-ray spectra for the \texttt{iD28Q0} (top) and \texttt{wD28Q0} (bottom) simulations at the final time snapshot. As in \autoref{fig:timeseries_case}, black points show the summed emission from all processes, while colored points connected by lines show the contribution from each individual emission process, with the emission from primary and secondary electrons shown separately. Pink points with error bars show $\gamma$-ray measurements from the Fermi 4FGL-DR4 catalog \citep{fermiDR4}. Black open circles show radio observations as compared in \citet{roth23} while black crosses show observations described in \citet{For18a}.}
        \label{fig:spectra_case}
\end{figure*}

\begin{figure*}
        \centering
    	\includegraphics[width=0.98\linewidth]{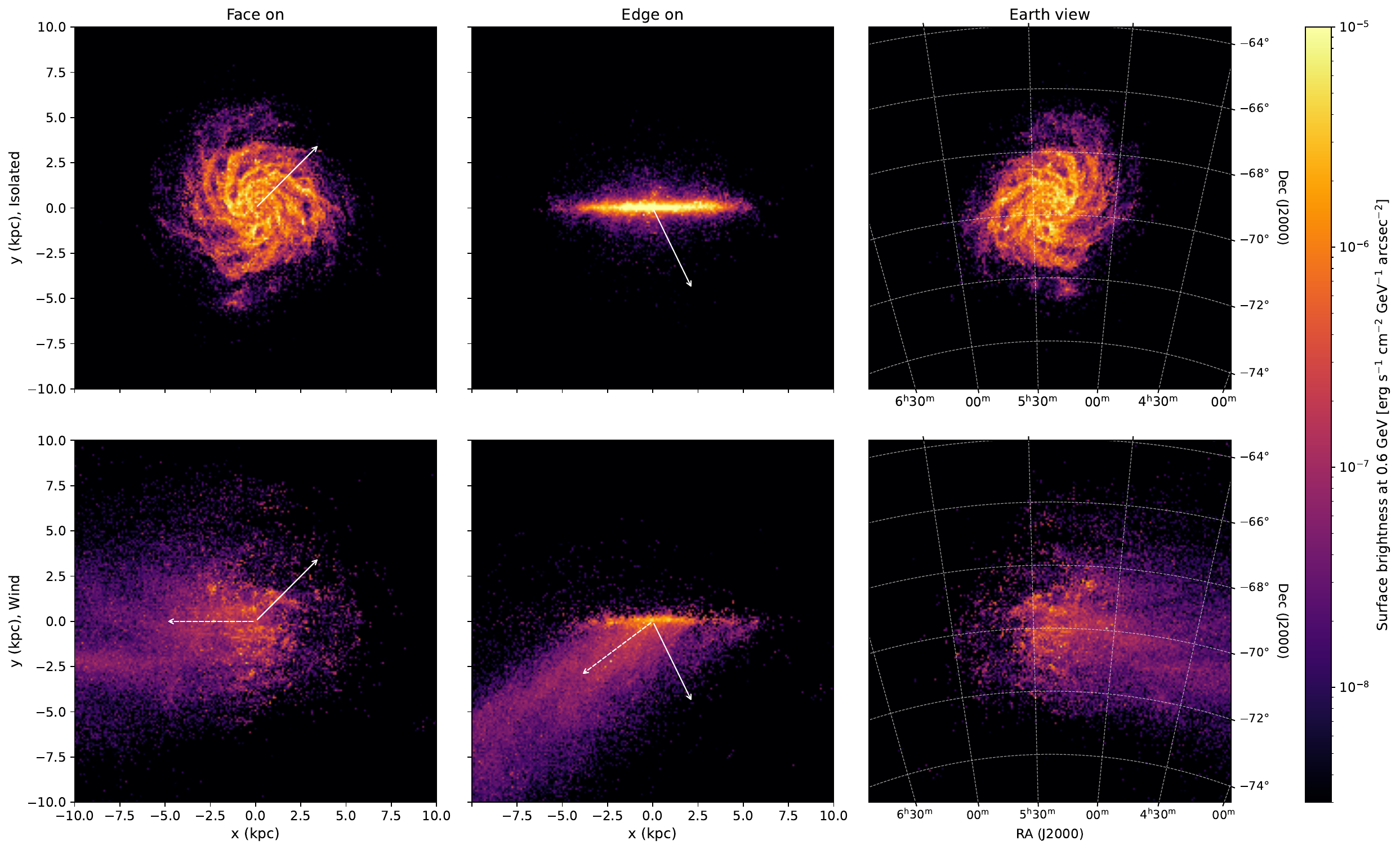}
    	\caption{Specific $\gamma$-ray flux at 0.6 GeV from the \texttt{iD28Q0} simulation (top row) and the \texttt{wD28Q0} simulation (bottom row). The three columns show three different projections of the LMC disk: face-on (left), edge-on (middle), and as viewed from Earth (right). The white dashed arrows indicate the direction of the wind as seen by an observer comoving with the LMC. The white solid arrows indicate the direction of Earth. For the purposes of assigning sky positions, we take the LMC's center to be at RA 05$^\mathrm{h}$ 23$^\mathrm{m}$ 34.6$^\mathrm{s}$ and Dec $-69^\circ45.4'$ \citep{tully13}, and we have rotated the image so that the direction of the LMC's proper motion in the wind simulation is oriented correctly on the sky---see \citetalias{shah25} for details.\\}
        \label{fig:gamma_map_onecase}
\end{figure*}

We use simulations \texttt{iD28Q0} and \texttt{wD28Q0} for our example, and for these two runs we begin our analysis by plotting the total (energy-integrated) $\gamma$-ray flux as a function of time in \autoref{fig:timeseries_case}, where we compute the total flux as
\begin{equation}
    F_\gamma = \frac{1}{4\pi d^2}\sum_i\left(\varepsilon_{i+1} - \varepsilon_i\right) \left(\frac{L_{\varepsilon,i+1} + L_{\varepsilon_,i}}{2}\right).
\end{equation}
Here $\varepsilon_i$ is the $i$th photon energy at which we compute $\gamma$-ray emission, $L_{\varepsilon,i}$ is the specific luminosity at this energy computed by \criptic, and $d = 49.6$ kpc is the distance to the LMC \citep{Pietrzynski19a}; our expression corresponds to a trapezoidal rule integration of the spectrum computed in \autoref{subsec:spectrum}. We carry out this integration both for the total $\gamma$-ray flux summed over all emission processes and process-by-process, and for the observed $\gamma$-ray data from the Fermi 4FGL-DR4 catalog \citep{fermiDR4}. We also carry out an analogous calculation for the energy-integrated radio emission, integrating over the bounds of the radio data we use from \citet{For18a} and \citet{roth23}. We do not show the integrated flux of the observed radio data, as we do not expect our simulated synchrotron flux alone to match the observed data without a separately modeled free-free flux.

We see that by 40 Myr of evolution the total flux and the flux produced by each individual process have mostly reached approximate steady-state; the primary exceptions are synchrotron, which is still rising for the reasons discussed in \autoref{subsec:spectrum}, and transient spikes in emission, primarily via inverse Compton. These spikes occur when supernovae inject large numbers of CRs near the LMC center, where the radiation field is particularly intense. We also see that the integrated emission is roughly an order of magnitude higher in the isolated case than in the wind case. The isolated case over-predicts the observed flux by about an order of magnitude, while the wind case is close to the observations.

We show the full emission spectra at the final time snapshot for both simulations in \autoref{fig:spectra_case}. From this figure we can make a few immediate observations. First, consistent with what we saw for the integrated $\gamma$-ray luminosity, the isolated simulation significantly overshoots the observed spectrum at all $\gamma$-ray energies, while the wind simulation produces a spectrum that is very close to the observed one at lower energies, with an an overshoot by a factor of a few at higher energies. Second, $\gamma$-ray emission is dominated by neutral pion decay in the isolated simulation but is more evenly split by process in the wind simulation, with inverse Compton, bremsstrahlung, and pion decay all contributing at the same order of magnitude. The integrated neutral pion decay and bremsstrahlung emission in the wind case is a factor of $\sim 10$ lower than the isolated case whereas the integrated inverse Compton emission is similar between the cases, leaving inverse Compton as the single largest contributor, albeit only by a factor of $\approx 2$. Similarly, radio synchrotron emission is a factor of $\approx 10$ times dimmer in \texttt{wD28Q0} than in \texttt{iD28Q0}, though in both cases the radio luminosity is close to observed values because free-free dominates except at the lowest energies. While \texttt{iD28Q0} appears to provide a near exact match to the observed radio data, we emphasize that this is likely a coincidence, as our simulated synchrotron emission has likely not equilibrated yet.

To understand the origin of these differences in the non-thermal emission, it is helpful to visualize the location of $\gamma$-ray emission, which we show in \autoref{fig:gamma_map_onecase}. From this figure it is clear that for this particular transport model adding a wind to the LMC pushes cosmic ray emission away from the LMC disk, which creates a dim tail in the direction of the wind in addition to significantly reducing the total radio and $\gamma$-ray luminosities. In particular, the central regions of the galaxy are significantly dimmer in \texttt{wD28Q0} than \texttt{iD28Q0}. This Figure therefore suggests an obvious interpretation for the reduced luminosity in \texttt{wD28Q0}: in the simulation including a wind, CRs that escape the LMC disk even by a small distance are likely to be advected away from the galaxy by the wind. This greatly reduces the amount of time that these CRs spend in the dense parts of the galaxy where collisional emission process such as pion production and bremsstrahlung are efficient, and where magnetic fields are strong and thus synchrotron emission is efficient. By contrast, the radiation field falls off with distance from the galaxy much more slowly than the density or magnetic field strength, and so the reduction in inverse Compton emission is much milder. The net effect is to make inverse Compton emission the most important process in the wind case, and to lower the total $\gamma$-ray luminosity by an order of magnitude.

\subsection{Results for varying CR transport prescriptions}
\label{subsec:diffusion_on_emission}

\begin{figure*}
        \centering
    	\includegraphics[width=\linewidth]{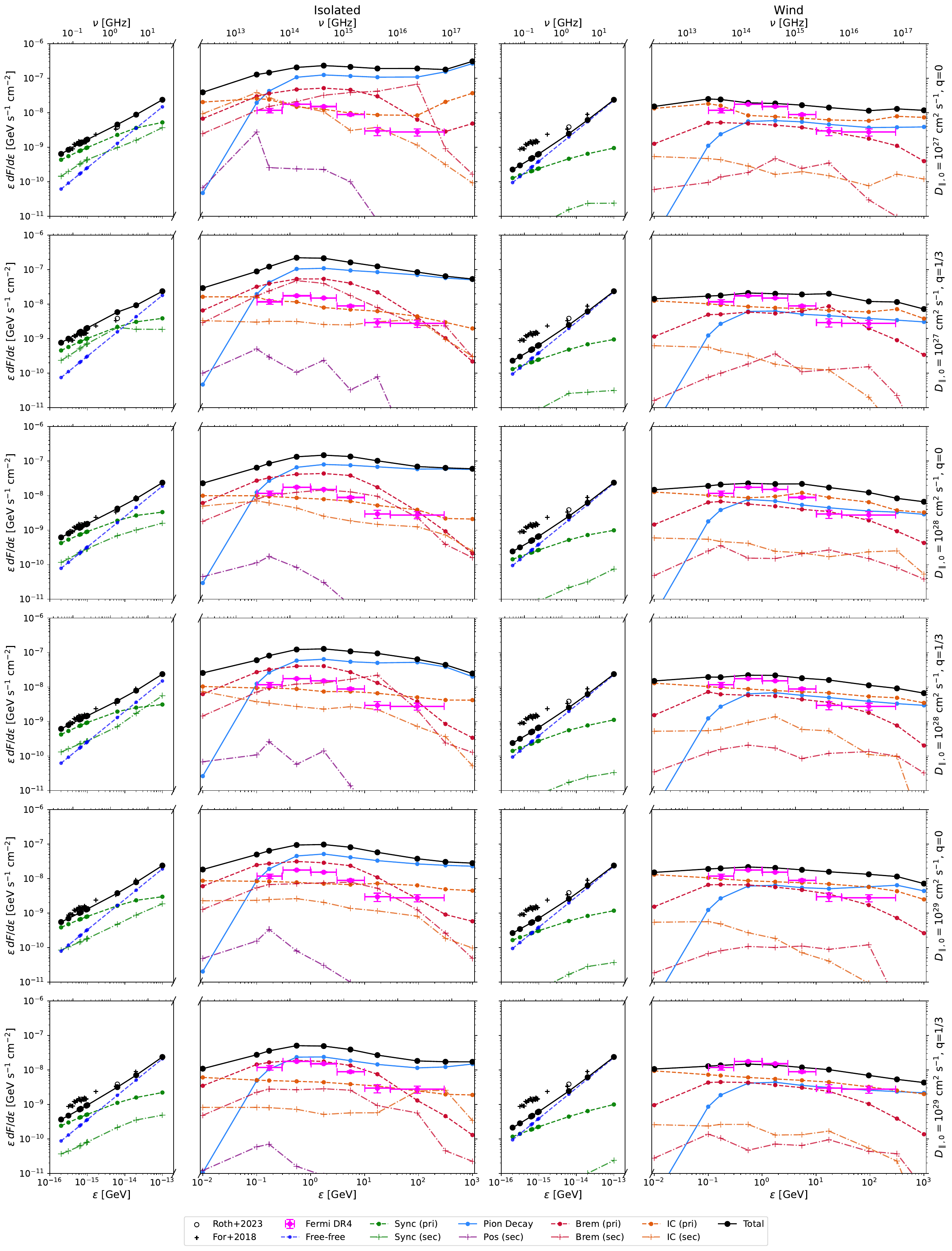}
    	\caption{Same as \autoref{fig:spectra_case}, but now showing the results for all simulations. The left column shows isolated simulations and the right column shows wind simulations. The value of the parallel diffusion coefficient in each simulation is indicated to the right of each row.}
        \label{fig:spectra_big}
\end{figure*}

\autoref{fig:spectra_big} shows the final emission spectra for all simulations (again separated by emission process), to illustrate the effect of different choices of diffusion coefficients on emission. For all diffusion coefficients, the wind simulations' emission in both radio and $\gamma$-rays is lower than the matching isolated simulation. Moreover, we see that $\gamma$-ray emission in the isolated simulations decreases as we increase the diffusion coefficient, while emission in the wind simulations stays roughly constant. The only noticeable difference between any of the wind simulations is a small decrease in $\gamma$-ray emission for the \texttt{wD29Q1/3} case. This is consistent with our interpretation that advection in the wind is the dominant pathway to CR escape in the wind simulations, contrary to the standard assumption that diffusive escape dominates in quiescent (i.e., non-starburst) galaxies. The small difference between the \texttt{wD29Q0} and \texttt{wD29Q1/3} cases could be a hint that diffusive escape is beginning to be significant for this very high diffusion coefficient; however, we emphasize this diffusion coefficient is already higher than previous studies, which have generally arrived at values of order $10^{28}$ cm$^2$ s$^{-1}$ \citep[e.g.,][]{werhahn21b, reynoso25}, an order of magnitude smaller. By contrast, the isolated simulations follow the standard expectation for models where diffusive transport dominates: larger diffusion coefficients lead to decreased emission because CRs escape more rapidly. However, even for the highest diffusion coefficient we have tested, the isolated simulation still overshoots the observed $\gamma$-ray flux by a factor of $\approx 5-10$ at all energies.

The similar emission spectra seen in all diffusive wind simulations means that the conclusions drawn from the \texttt{D28Q0} pair of simulations generally apply to all diffusive simulations. For all wind simulations we find that inverse Compton emission is the most significant source of emission, with neutral pion decay and bremsstrahlung as secondary contributors. We also find that the $\gamma$-ray spectrum provides a good match to the observed data below $\sim 10$ GeV, but overestimates the observed spectrum at higher energies by a factor of a few. One possible explanation could be that the diffusion coefficient has a stronger energy dependence than the $D_\parallel \propto p^{1/3}$ model that we test. However, this possibility would require extremely high diffusion coefficients for CRs with $\sim \mathrm{TeV}$ energies: our highest diffusion simulation already has a diffusion coefficient of $10^{30}$ cm$^2$ s$^{-1}$ at an energy of 1 TeV (see \autoref{eq:D_powerlaw}), so suppressing the emission via diffusive escape further would require even larger diffusion coefficients. Such a high diffusion coefficient is not generally considered in standard ISM environments, but the effect of the wind could potentially push up the diffusion coefficients within the LMC.

A second possibility is that our assumed injection spectrum with slope $-2.2$ out to PeV energies is not steep enough at high energies, meaning we inject too many CRs with high energies. Adding slight steepening in the injection spectrum above $\sim 1$ TeV, or considering a range of cutoff energies with only some supernova remnants driving particles to $\gg\mathrm{TeV}$ energies, might resolve the remaining tension with the observations.

The difference in the synchrotron spectra within all models in figure \ref{fig:spectra_big} is similar to the difference between the $\gamma$-ray spectra: a larger diffusion coefficient reduces synchrotron emission in the isolated case while having little effect in the wind cases, indicating that diffusive escape dominates for isolated simulations but advective escape dominates for wind ones. Similarly, in all wind cases, the synchrotron emission underestimates the radio emission of the LMC at frequencies below $\sim 1$ GHz assuming our free-free model is correct. However, as discussed in \ref{subsec:spectrum}, this is likely because the loss time for the CR electrons responsible for driving radio emission at low frequencies is longer than our simulation time, and thus the low-frequency synchrotron luminosity has not yet reached steady-state even at our final snapshot. 

\section{Discussion}
\label{sec:discussion}

Our goal in this discussion is to understand in more detail why simulations including the Milky Way wind seem able to reproduce the observed $\gamma$-ray luminosity of the LMC while those omitting it do not, and to consider the observational implications of these results. To this end, we first in \autoref{subsec:adiabatic} examine the loss processes affecting CR protons in our simulations, and in \autoref{subsec:calorimetry} use the results of \autoref{subsec:adiabatic} to quantify the degree of proton calorimetry in our simulations, and how the presence of a wind affects it. Finally in \autoref{subsec:observations} we discuss possible observational tests of our results.

\subsection{Loss processes for CR protons}
\label{subsec:adiabatic}

\begin{figure*}
        \centering
    	\includegraphics[width=0.98\linewidth]{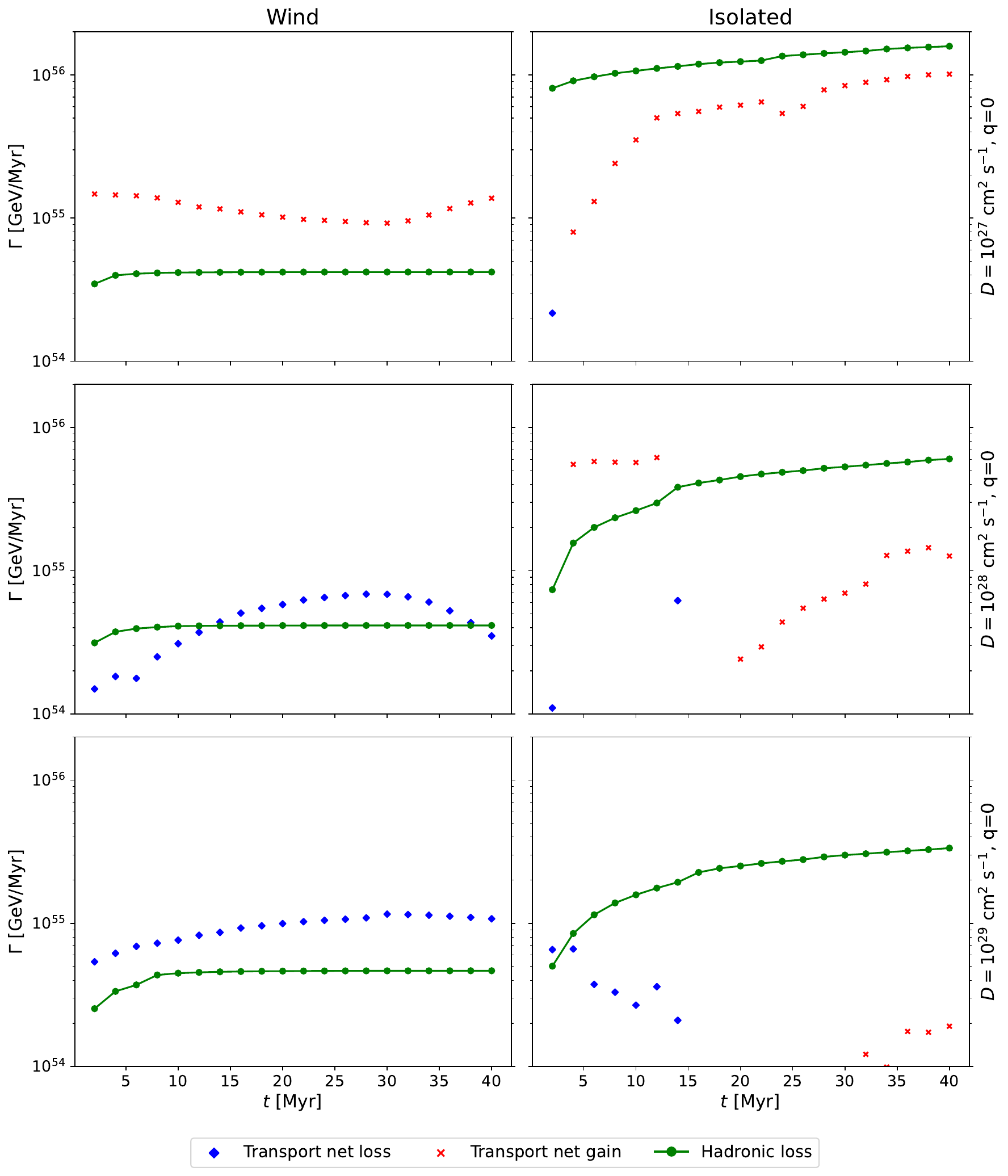}
    	\caption{Total energy loss rates ($\Gamma$) due to proton-proton collisions and adiabatic work for all proton packets within the $q=0$ simulations. Each point sums up all particles that are younger than the $t$-axis age, so the rightmost points sum up all particles within the simulation. Each point is averaged over 2.5 Myr of simulation time.}
        \label{fig:transport_loss}
\end{figure*}

It is helpful to begin this analysis by examining the processes that affect the energy content of an individual CR proton packet, representing a population of protons all with the same kinetic energy and position. Each such packet suffers losses from both transport and microphysical processes. For CR protons in the energy range that contribute to the emission detected by \textit{Fermi} (roughly $\gtrsim 1$ GeV), the dominant microphysical loss process is proton-proton collisions (i.e., hadronic losses); Coulomb and ionization losses occur as well, and are included in \criptic, but these are strongly sub-dominant for the energies of interest to us, and so we will neglect them for simplicity. In addition to these microphysical processes, CRs also undergo losses (or gains) due to adiabatic work that the gas does on the CR field. It is therefore of interest to quantify the relative importance of transport versus microphysical losses.

For $pp$ collision losses, we define the rate at which a proton packet with a statistical weight $w_i$ (i.e., which represents $w_i$ individual protons) and kinetic energy $T_i$ loses energy as (c.f.~equation 17 of \citetalias{criptic})
\begin{equation}
    \Gamma_{pp,i} = w_i T_i \eta_\mathrm{in} \sigma_{\mathrm{nuc},i} v_i \frac{\rho}{\mu_\mathrm{H} m_\mathrm{H}},
\end{equation}
where here $v_i$ and $\sigma_{\mathrm{nuc},i}$ are the velocity and nuclear inelastic scattering cross section per H nucleon for a proton of kinetic energy $T_i$, $\eta_\mathrm{in} \approx 1/2$ is the fraction of a proton's initial energy lost per $pp$ interaction, $\rho$ is the ambient gas density at the packet's position, $\mu_\mathrm{H} \approx 1.4$ is the mean mass per H nucleon for Milky Way elemental abundances, and $m_\mathrm{H}$ is the hydrogen mass. For adiabatic losses (or gains), packets change their momenta at a rate $\dot{p} = -(p/3) \nabla\cdot \mathbf{v}_g$, where $\mathbf{v}_g$ is the velocity of the gas at the CR position, and thus the corresponding rate of adiabatic loss is
\begin{equation}
    \Gamma_{\mathrm{adb},i} = \frac{1}{3} p_i \left(\frac{dT}{dp}\right)_i \nabla \cdot \mathbf{v}_\mathrm{g},
\end{equation}
where $p_i$ is the packet momentum and $(dT/dp)_i$ is the derivative of kinetic energy with respect to momentum at momentum $p_i$. Note that the sign convention we have adopted here is that positive values indicate a loss of CR energy to the gas, which occurs when $\nabla\cdot\mathbf{v}_g > 0$, while negative values indicate a gain of CR energy from the gas.

To examine how these loss rates behave in our simulations, we compute $\Gamma_{pp,i}$ and $\Gamma_{\mathrm{adb},i}$ for all packets at the final snapshot in each simulation. We plot the summed energy loss rate for all CR packets younger than age $t$ (where the packet age is defined as the time since it was injected) as a function of $t$ in \autoref{fig:transport_loss}; here we show only the $q=0$ simulations for brevity, but the results are qualitatively identical for the $q=1/3$ cases. Since we are plotting a cumulative sum in age, small values on the horizontal axis illustrate results for only younger packets, while the rightmost data point at $t=40$ Myr sums over all packets in the simulations and thus indicates which processes dominate in steady-state.

First examining the hadronic loss curves (green lines in \autoref{fig:transport_loss}), we see hadronic losses are smaller overall in the wind cases than in their corresponding isolated runs, and that in wind cases the majority of hadronic losses occur shortly after injection, with packets older than $\sim10$ Myr experiencing negligible additional loss and therefore contributing very little to pion decay emission. In the isolated simulations, by contrast, the loss rate drops off with age much more gradually, with packets injected at the start of the simulation (i.e., with an age of 40 Myr) still losing energy to hadronic collisions, which contributes to $\gamma$-ray emission. This is consistent with the suggestion that CRs escape from the dense regions of the galaxy much more rapidly as a result of the Milky Way wind.

Interestingly, however, there is also a  significant difference in adiabatic losses in the wind cases versus the isolated ones. For \texttt{wD28Q0} and \texttt{wD29Q0}, in steady state (i.e., for older ages) more energy is lost to adiabatic expansion than to $pp$ collisions, and these adiabatic losses are stronger than in the isolated case, whereas in all the isolated runs at almost all times, adiabatic gains and losses are subdominant compared to $pp$ losses. This is another reason why pion-decay emission is lower in the wind cases, and has a clear physical explanation: the ``tail'' of material stripped from the LMC by the Milky Way wind (right panel of \autoref{fig:lmc_sim}) is under-pressured compared to the Milky Way CGM in the transverse direction, and so the stripped material undergoes rapid adiabatic expansion. Any CRs entrained with this material will therefore suffer strong adiabatic expansion losses.

By contrast, in the lowest diffusion rate cases, \texttt{wD27Q0} and \texttt{iD27Q0}, adiabatic gains outweigh adiabatic losses. In the isolated case adiabatic gains are subdominant compared to $pp$ losses, and thus are not energetically significant, but in the wind case adiabatic gains actually outweigh $pp$ losses. This pattern arises because adiabatic gains occur in CR protons trapped inside collapsing or compressing regions of gas from which they cannot escape---a situation that is more likely when the diffusion coefficient is smaller. However, there is an important caveat to this discussion, which is that \criptic~does not model the back-pressure that the cosmic ray field applies to the collapsing gas, which might limit compression. This means the adiabatic gains may be overestimated. Nonetheless, the cases where this is a concern use an extremely low value for the diffusion coefficient. It is more likely that the higher diffusion cases are closer to reality, in which case we have the interesting result that adiabatic losses are comparable to or more important than $pp$ losses in regulating the energy balance of CR protons in the LMC; this is in strong contrast to the conventional assumption that $pp$ losses are dominant.

Finally, we pause to note one additional feature in the wind simulations: in all these runs, the adiabatic loss rate suddenly decreases, or the gain rate increases, when including particles older than $\sim30$ Myr, meaning particles of that age experience adiabatic gains in all runs. This gain occurs when CRs have had enough time to reach the bow shock of the LMC, visible in \autoref{fig:lmc_sim}. This shock compresses the fluid, leading to the adiabatic gains, though there is little corresponding effect on $\gamma$-ray emission because the region in which this occurs is too diffuse to generate significant $pp$ losses. While \criptic~does not treat diffusive shock acceleration of CRs, the fact that adiabatic gains in the shock region are significant raises the intriguing possibility that shock acceleration might also occur, which could potentially create a population of ultrahigh energy CRs (UHECRs) outside the disk of the LMC, which would affect the CR population in the nearby CGM. Observations of EeV UHECR events by the Pierre Auger Observatory suggest that these events are clustered towards the Centaurus A region of the sky \citep{auger25}, and \citet{clay25} propose that the LMC---which is close to Centaurus A in the sky---could also be a source of UHECRs. They propose the 30 Doradus region as a likely acceleration site, but our findings suggest that acceleration at the LMC bow shock is another possibility.

\subsection{Proton calorimetry}
\label{subsec:calorimetry}

The presence of strong adiabatic gains and losses in our wind simulations motivates a careful definition of the proton calorimetry fraction, which is typically defined as the proportion of CR proton energy injected by supernovae that is lost to hadronic collisions. In the absence of strong adiabatic gains and losses this can be calculated by tracking the energy of individual CR proton packets over time and comparing their equilibrium energy content with their energy at injection, but this approach breaks down when CRs can gain or lose significant amounts of energy adiabatically over extended periods of time. We therefore adopt an alternative approach: we define two approximate instantaneous calorimetric fractions: one that ignores adiabatic effects and is defined as
\begin{equation} 
\label{eq:eta_bare}
\eta = \frac{\Gamma_{pp}}{\langle L_\mathrm{src} \rangle}
\end{equation}
and a second that is corrected for adiabatic effects and is defined as
\begin{equation} 
\label{eq:eta_adb}
\eta_\mathrm{adb} = \frac{\Gamma_{pp}}{\langle L_\mathrm{src} \rangle - \Gamma_\mathrm{adb}}.
\end{equation}
In these expressions, $\Gamma_{pp}$ and $\Gamma_\mathrm{adb}$ are the total $pp$ and adiabatic energy loss rates summed over all proton packets at the final snapshot of our simulations, and $\langle L_\mathrm{src} \rangle$ is the time-averaged CR proton energy injection rate over the course of a given simulation; this is $\langle L_\mathrm{src} \rangle= 5.68\times 10^{55}$ GeV/Myr for the isolated cases and $\langle L_\mathrm{src} \rangle = 6.91\times 10^{55}$ GeV/Myr for the wind cases. Note that both $\Gamma_\mathrm{pp}$ and $\Gamma_\mathrm{adb}$ are defined as positive when there is a net loss, so $\Gamma_\mathrm{pp}$ is always positive and $\eta_\mathrm{adb}$ will be less than $\eta$ when there are strong adiabatic net gains.

Before proceeding with calculations of $\eta$ and $\eta_\mathrm{adb}$, it is important to note some caveats. First, we are using the time-averaged CR luminosity in the denominator, but in reality the simulation star formation rates, and thus supernova rates, vary by factors of $\sim 1.5-2$ over $\sim 10$ Myr timescales \citep[c.f.~Figure 4 of ][]{shah25}; since younger packets contribute disproportionately to $pp$ losses and thus $\gamma$-ray production, this means we should expect fluctuations in our computed calorimetry fractions at this level. Moreover, our definitions of the calorimetric fraction here implicitly assume that $40$ Myr old packets represent a population that has reached statistical steady state; this assumption is likely true in all wind cases, but is more questionable in the isolated cases where $pp$ losses are still occurring even for the oldest packets. These limitations mean that the calorimetric fractions represent a useful way of comparing different simulations, but should be treated with caution for the purposes of comparing with other models or measurements.

\begin{table}
    \centering
    \begin{tabular}{lccclcc}
    \hline\hline
    \\[-2ex]
    \multicolumn{3}{c}{Isolated simulations} & \qquad &
    \multicolumn{3}{c}{Wind simulations} \\
    Simulation & $\eta$ & $\eta_\mathrm{adb}$ & &
    Simulation & $\eta$ & $\eta_\mathrm{adb}$ 
    \\
    \\[-2ex]
    \hline
    \\[-2ex]
    \texttt{iD27Q0}    & 2.789 & 1.003 & &
    \texttt{wD27Q0}    & 0.061 & 0.051\\
    \texttt{iD27Q1/3}  & 1.784 & 0.576 & &
    \texttt{wD27Q1/3}  & 0.062 & 0.067\\ 
    \texttt{iD28Q0}    & 1.065 & 0.871 & &
    \texttt{wD28Q0}    & 0.060 & 0.063\\ 
    \texttt{iD28Q1/3}  & 1.034 & 0.834 & &
    \texttt{wD28Q1/3}  & 0.066 & 0.073\\ 
    \texttt{iD29Q0}    & 0.589 & 0.570 & &
    \texttt{wD29Q0}    & 0.067 & 0.080\\
    \texttt{iD29Q1/3}  & 0.294 & 0.331 & &
    \texttt{wD29Q1/3}  & 0.050 & 0.051\\
        \hline\hline
        \end{tabular}
        \caption{Calorimetric fractions from all simulations, using the definitions in \autoref{eq:eta_bare} and \autoref{eq:eta_adb}. 
        \label{tab:calfrac}
        }
    \end{table}

\autoref{tab:calfrac} summarizes the calorimetric fractions calculated from all simulations. Several of the isolated simulations have a calculated fraction greater than unity, which is possible in our models because of adiabatic gains and the variable energy injection rate. The wind cases consistently have lower fractions than the isolated simulations, corresponding with their lower simulated emission. For the isolated cases, higher diffusion corresponds with lower $\eta$ and generally lower $\eta_\mathrm{adb}$--except for the \texttt{iD27Q1/3} case, which has a particularly high total adiabatic energy gain rate. Conversely, the wind simulations shows small deviations between the diffusion cases, except for the \texttt{wD29Q1/3} case with a slightly lower fraction under both definitions. This is consistent with the wind cases having similar emission for all cases with slightly lower emission in the highest diffusion case. The wind simulations also show relatively little difference between $\eta$ and $\eta_\mathrm{adb}$, simply because for all the wind simulations $\Gamma_\mathrm{adb} \ll \langle L_\mathrm{src}\rangle$, i.e., of the injected energy only a small fraction is lost to \textit{either} $pp$ collisions or adiabatic expansion in the wind cases.

Thus the main conclusion we can draw from this section, in conjunction with the previous one, is that, while adiabatic losses can in fact exceed $pp$ losses for wind runs, this is a relatively small part of the explanation for why our wind runs do a better job of reproducing the observed LMC luminosity. The main effect is simply that, due to advective loss into the wind, only a small fraction of the injected proton energy is lost to \textit{either} $pp$ collisions or adiabatic expansion. Most of the CR proton energy simply escapes the galaxy by advection, leaving the calorimetry fraction well below the $\sim 10-20\%$ that is conventionally assumed for non-starburst galaxies.

\subsection{Observational tests}
\label{subsec:observations}

\begin{figure*}
        \centering
    	\includegraphics[width=0.98\linewidth]{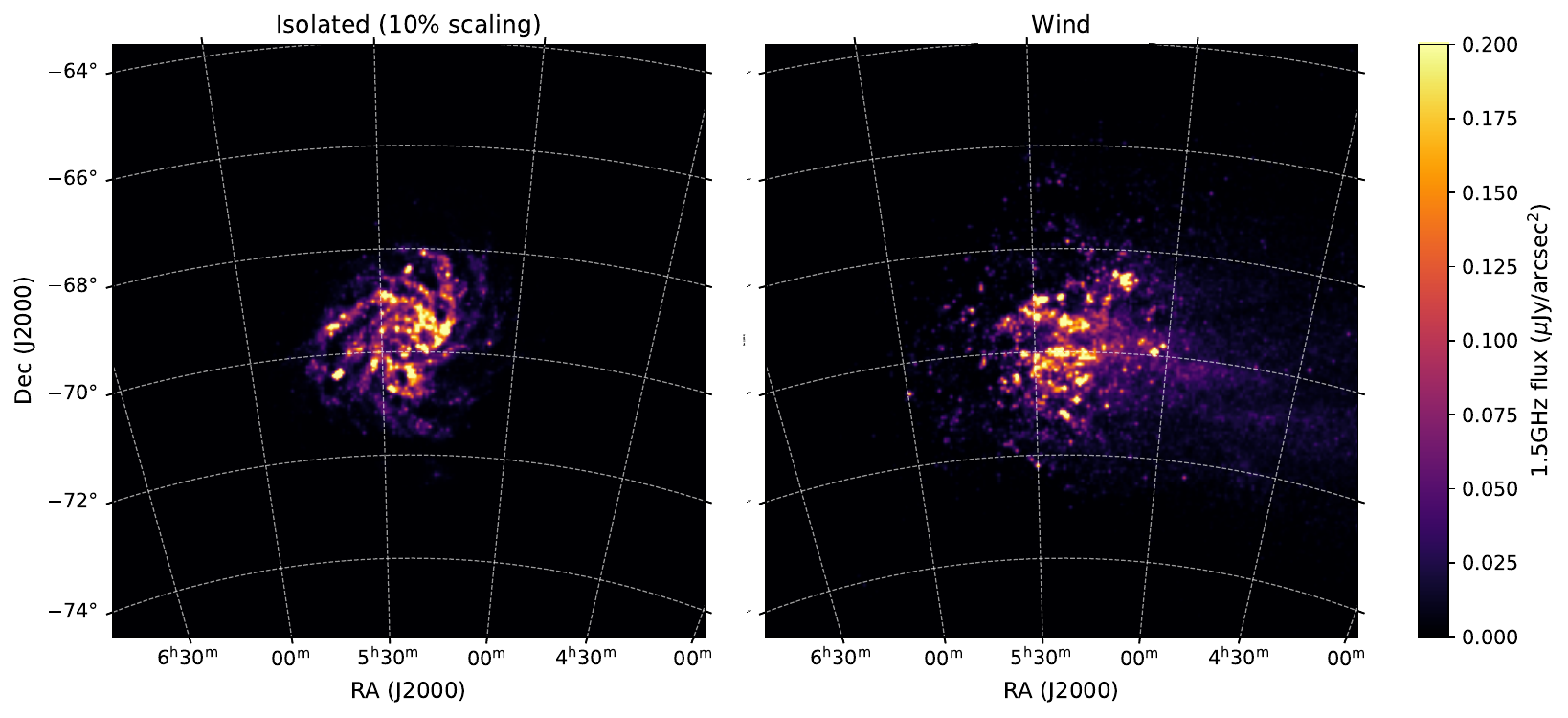}
    	\caption{Predicted surface brightness at 1.5 GHz for the \texttt{iD28Q0} and \texttt{wD28Q0} cases as viewed from Earth. The field of view shown is 20 kpc $\times$ 20 kpc at the distance to the LMC. In order to generate this figure we have convolved the true surface brightness with a Gaussian kernel with with $\sigma=0.5$ kpc, comparable to the expected resolution of single dish telescopes at LMC distances. For the isolated case we have scaled down the predicted surface brightness by a factor of 10 so that we can show it one the same colorbar as the wind case. \\}
        \label{fig:radio_map}
\end{figure*}

Given our finding that the LMC's low $\gamma$-ray luminosity is primarily due to advective escape of CRs into the tail of material stripped by the LMC's interaction with the Milky Way, it is interesting to ask where there are any potentially observable signatures of this effect that could serve as independent tests of our finding. One immediate candidate is the morphology of the $\gamma$-ray emission: \autoref{fig:gamma_map_onecase} shows the predicted $\gamma$-ray emission for the \texttt{iD28Q0} and \texttt{wD28Q0} cases, and reveals that the wind case features a tail of faint $\gamma$-ray emission that is absent in the isolated case. However, given the low angular resolution and surface brightness sensitivity of $\gamma$-ray observatories, it is unlikely that this faint tail will be detectable in the foreseeable future: even in our figures it is only visible because we are plotting emission on a logarithmic scale, and would be invisible on a linear scale, since the tail is fainter than the main body of the galaxy by a factor of $\sim 10^2 - 10^3$.

Given that we cannot feasibly observe the $\gamma$-ray tail our simulations predict, we focus our attention on radio emission as a more promising possibility. Radio has two main advantages. One is that radio telescopes are substantially more sensitive and high-resolution than $\gamma$-ray ones, and the other is that we expect the the tail to be brighter relative to emission from the main body of the LMC in radio than in $\gamma$-rays because the low density of the tail ensures that there is negligible $pp$ emission from it (the $\gamma$-ray emission that is visible is almost entirely inverse Compton), but the magnetic field strength in the tail is only weaker than that in the galactic disk by a factor of a few (c.f.~Figure 10 of \citetalias{shah25}). Given these considerations, in \autoref{fig:radio_map} we show the predicted emission from the isolated and wind simulations at 1.5 GHz. We choose this frequency to balance two competing effects: lower frequency emission is less likely to have equilibrated within our simulations due to the long cooling times of electrons whereas higher frequency emission is dominated by free-free emission that \criptic~does not model (see \autoref{subsec:spectrum} and \autoref{eq:cooling_time}). Our estimated cooling time at 1.5 GHz is $\sim$66 Myr, which is still longer than our simulation run time, but by less than a factor of two. Thus we may regard the predicted synchrotron emission as a lower limit, but a relatively good one. \autoref{fig:radio_map} reveals that a faint tail in the wind simulation is indeed visible even when the data are shown on a linear scale. The brightness of this tail compared to the emission from the disk is reduced by roughly a factor of 10, in contrast with the $\gamma$-ray maps where the contrast is close to a factor of $10^2-10^3$. 

While the tail is more visible compared to the rest of the disk in radio than in $\gamma$-ray, this does not guarantee that it is detectable. Indeed, the predicted surface brightness of order $0.05$ $\mu$Jy arcsec$^2$ at 1.5 GHz is likely beyond easy reach for existing facilities. For example, ASKAP has a beam of a few hundred arcsec$^2$ at frequencies near 1 GHz, so the brightness of the tail would correspond to $\approx 1-10$ $\mu$Jy per beam, while a 10-hour integration in ASKAP only reaches a noise level of $\approx 20$ $\mu$Jy per beam, meaning that detecting the tail would require hundreds of hours of integration. Moreover, this ignores the potentially serious issue of confusion with the Milky Way foreground, which is likely to be more limiting than telescope noise at these low surface brightnesses. One can conceivably overcome this problem using template-based analysis or using information about the spectral shape of the tail emission, but modeling such strategies is beyond the scope of this work. Thus while the existence of the faint radio tail is a potential test of the model we present, in practice it may be an extremely difficult test to perform.

\subsection{The SMC}
Our conclusions on the effect of ram-pressure stripping on the LMC's $\gamma$-ray emission should in principle apply to the SMC too. The SMC also experiences ram-pressure stripping due both to interactions with the Milky Way CGM and tidal forces from the LMC \citep[e.g.][]{nakano_25}, and thus one might expect the SMC's $\gamma$-ray luminosity to be overestimated in models that do not include a wind just as the LMC's is. Contrary to this naive expectation, however, the SMC fits cleanly on the same FIR$\gamma$ relation as galaxies other than the LMC \citep{kornecki20, werhahn21b}. 

In part this may be a data artifact. As noted by \citep{roth23}, published estimates for the SMC's $\gamma$-ray spectrum differ by substantially more than the reported error bars, likely due to systematic uncertainties associated with separating it from the Milky Way foreground. However, the difference with the LMC could also be due to complex dynamical effects unique to the SMC, for example a different angle of attack between the galactic disk and the oncoming wind from the Milky Way CGM, or the fact that the SMC is losing gas both due to the CGM wind and due to tidal interactions with the LMC. Investigating this further would require a simulation of the SMC that is calibrated against magnetic field data---such that CRs are accurately modeled---and that takes into account the dynamical effects of both the Milky Way's CGM and the LMC.

Finally, we note that the data do contain a suggestive hint that the advective loss effect we have identified for the LMC may also operate in the SMC. While \citet{roth23} obtain a relatively good prediction for the SMC's total luminosity, the $\gamma$-ray spectrum that they predict overestimates the $\gamma$-ray flux at $\sim$1 GeV, and in this regime, the two published spectra for the SMC agree \citep{ajello20, fermi43}. (The disagreement is at higher energies.) On the other hand, the observed data at $\sim1$ GeV roughly match the inverse Compton spectrum predicted by \citealt{roth23}. This is a potential hint that a wind in the SMC's frame may also reduce its pion decay emission, explaining the reduced emission at $\sim 1$ GeV. 

\section{Conclusion}
In this work, we use \criptic~to model cosmic ray driven emission in the LMC to explain its surprising $\gamma$-ray dimness. We consider six different models for diffusion-based CR transport on top of two simulations of the LMC---one where the LMC experiences a headwind due to its interaction with the Milky Way's CGM, and a control run with identical initial conditions but without the CGM interaction. We find that the headwind the LMC experiences as it falls into the Milky Way CGM advects cosmic rays away from the LMC disk and therefore significantly reduces its $\gamma$-ray emission, independent of the CR transport model we adopt. This manifests in a reduced $\gamma$-ray luminosity, with the main source of emission becoming leptonic inverse Compton emission rather than hadronic pion decay emission. As a result of this effect, all our simulations including the LMC's CGM interaction yield $\gamma$-ray luminosities in good agreement with observations, while the corresponding control runs predict higher-than-observed $\gamma$-ray luminosities, reproducing a problem found in many previous models.

The $\gamma$-ray spectrum in our wind simulations is largely independent of the chosen diffusion coefficient, supporting advective escape as the dominant CR loss mechanism. Adiabatic losses also become significant for some of our wind simulations, but are subdominant compared to advective loss as an explanation for the overall spectral shape and intensity. The $\gamma$-ray spectrum we model from all wind simulations is a close match to the observed spectrum at photon energies below $\sim 10$ GeV, but significantly overestimates the observed spectrum at higher energies. This discrepancy could be explained by a larger diffusion coefficient, a steeper increase in diffusion coefficient with energy in the CGM, or a steeper CR injection spectrum than assumed in our models.

Finally, we find that the wind also forms a dim tail extending out of the LMC with brightness a factor of $10^2-10^3$ lower than the disk in $\gamma$-rays, and a factor of $10$ lower than the disk at $\sim 1$ GHz. While we investigate the possibility of directly observing this tail with radio telescopes, our predicted surface brightness is too low to be reliably observed with current and upcoming telescopes, and would likely be contaminated by Milky Way foreground emission. Nonetheless, if such a tail does become detectable thanks to next generation facilities such as the SKA, a search for it would represent a strong test of our model.

\section*{Acknowledgments}

MRK and RMC acknowledge support from the Australian Research Council through Discovery Projects awards DP230101055 and DP260100433. This research was undertaken with the assistance of resources from the National Computational Infrastructure (NCI Australia), an NCRIS enabled capability supported by the Australian Government, through award jh2.

\bibliographystyle{mnras}

\bibliography{oja_template}

\end{document}